# Dissecting Bias of ChatGPT in College Major Recommendations

Alex Zheng

Plano West Senior High School, alexzheng414@gmail.com

**Abstract**

Large language models (LLMs) such as ChatGPT play a crucial role in guiding critical decisions nowadays, such as in choosing a college major. Therefore, it is essential to assess the limitations of these models' recommendations and understand any potential biases that may mislead human decisions. In this study, I investigate bias in terms of GPT-3.5 Turbo's college major recommendations for students with various profiles, looking at demographic disparities in factors such as race, gender, and socioeconomic status, as well as educational disparities such as score percentiles. To conduct this analysis, I sourced public data for California seniors who have taken standardized tests like the California Standard Test (CAST) in 2023. By constructing prompts for the ChatGPT API, allowing the model to recommend majors based on high school student profiles, I evaluate bias using various metrics, including the Jaccard Coefficient, Wasserstein Metric, and STEM Disparity Score. The results of this study reveal a significant disparity in the set of recommended college majors, irrespective of the bias metric applied. Notably, the most pronounced disparities are observed for students who fall into minority categories, such as LGBTQ+, Hispanic, or the socioeconomically disadvantaged. Within these groups, ChatGPT demonstrates a lower likelihood of recommending STEM majors compared to a baseline scenario where these criteria are unspecified. For example, when employing the STEM Disparity Score metric, an LGBTQ+ student scoring at the $50^{th}$ percentile faces a 50% reduced chance of receiving a STEM major recommendation in comparison to a male student, with all other factors held constant. Additionally, an average Asian student is three times more likely to receive a STEM major recommendation than an African-American student. Meanwhile, students facing socioeconomic disadvantages have a 30% lower chance of being recommended a STEM major compared to their more privileged counterparts. These findings highlight the pressing need to acknowledge and rectify biases within language models, especially when they play a critical role in shaping personalized decisions. Addressing these disparities is essential to foster a more equitable educational and career environment for all students.

**Keywords:** Large Language Models (LLM), ChatGPT, Bias, Prompt Engineering, College Major Recommendation

# 1. Introduction

Large language models (LLMs) powered by artificial intelligence (AI), such as ChatGPT, have become integral to various aspects of our daily lives, from assisting with natural language understanding (Luo et al. 2023) to providing recommendations for critical decisions, including academic and career choices (Alwahaidi 2023). One prominent decision where these AI models play a crucial role is in providing college major recommendations for students (Stein et al. 2020, CollegeData 2023). These recommendations can greatly influence an individual's academic and career trajectory. Therefore, examining the potential biases present in these recommendations is of paramount importance to ensure fair and equitable guidance (Stein 2020).

This study delves into the critical issue of bias within LLMs concerning their college major recommendations for students from diverse backgrounds, including gender, race, socioeconomic status, and educational performance. It is essential to understand that the ability of LLMs to provide tailored recommendations for individuals can be a double-edged sword. On the one hand, it has the potential to offer highly personalized guidance that considers an individual's unique characteristics. On the other hand, it opens the door to the possibility of inadvertent or systemic biases infiltrating into these recommendations (Liang et al. 2021, Tsintzou et al. 2018). In this study's context, bias refers to the systematic and unfair discrimination by AI against certain individuals or groups, while favoring others in various aspects (Barocas et al. 2023). Failing to account for these biases may lead to suboptimal decisions at the individual level, perpetuating an inequitable societal system. Consequently, the importance of comprehending and identifying these biases cannot be overstated.

To investigate the presence of bias in LLM recommendations, I employ data from the year 2023, focusing on 12th-grade students who have taken standardized tests such as the California Standard Test (CAST). The choice of this dataset is motivated by its relevance to the high-stakes decision-making process that students face in their transition to higher education. I construct prompts for the ChatGPT API, which includes variables such as a student's academic performance, race, gender, and socioeconomic status, mirroring the data from my chosen dataset.

Bias in LLMs can manifest in various forms (Liang et al. 2021, Agarwal et al. 2022). It may result from a lack of representation in training data, inadvertent learning from historical biases present in text corpora, or structural issues in the architecture of the language model. Regardless of its origin, bias in recommendations can significantly impact the opportunities and choices available to individuals, perpetuating disparities in education and careers (Baker and Hawn 2021).

My analysis utilizes a comprehensive set of three metrics to evaluate the fairness and equity of recommendations (Pazzani and Billsus 2007, Tsintzou et al. 2018, Wang et al. 2013, Chen et al. 2023). The Jaccard coefficient measures the (dis)similarity of recommended majors between demographic groups (Pazzani and Billsus 2007). The STEM Disparity Score, built on the widely recognized fairness metric Disparate Impact (Barocas et al. 2023), assesses STEM major recommendation fairness (bias). Finally, the Wasserstein metric examines distributional differences in semantic similarity between recommendations (Chen et al. 2023).

The findings of this research reveal substantial disparities in the set of recommended majors, irrespective of the bias metric applied. Notably, the most pronounced disparities are observed for students who fall into minority categories, such as LGBTQ+, Hispanic, or the socioeconomically disadvantaged. Within these groups, ChatGPT demonstrates a lower likelihood of recommending STEM majors compared to a baseline scenario where these criteria are unspecified. For example, when employing the STEM Disparity Score metric, an LGBTQ+ student scoring at the $50^{th}$ percentile faces a 50% reduced chance of receiving a STEM major recommendation in comparison to a male student. Additionally, an average Asian student is three times more likely to receive a STEM major recommendation than an African-American student. Meanwhile, students facing socioeconomic disadvantages have a 30% lower chance of being recommended a STEM major compared to their more privileged counterparts.

This study underscores the necessity of recognizing and addressing bias in LLMs when making personalized decisions (West 2023). It is imperative to ensure that the recommendations provided by these models do not mislead students in their college major choices, regardless of their demographic backgrounds. Addressing these biases is essential to foster a more equitable educational and career environment for all students (Baker and Hawn 2021).

## 2. Literature Review

The intersection of AI and education has garnered significant attention in recent years, driven by the emergence of LLMs and their potential to assist students in making critical decisions. Alwahaidi (2023) highlights how students have begun using AI, particularly LLMs like ChatGPT, in their university applications. However, the integration of AI in education has raised concerns about algorithmic bias, particularly in academic contexts. Baker and Hawn (2021) emphasize the critical need to address and mitigate biases in AI systems used for educational purposes. Furthermore, the role and potential setbacks of ChatGPT in the college application process are discussed in CollegeData's resource, which underscores the importance of understanding AI's role in academic decisions and its potential impact on students' futures (CollegeData 2023). Stein et al. (2020) have delved into the development of college major recommendation systems, reflecting the practical applications of recommendation systems in guiding students through academic choices.

This study is also related to the nascent literature on machine bias/fairness. In an era where AI increasingly influences decision-making (Ning et al. 2023), fairness and machine learning have become pivotal topics of interest. Barocas et al. (2023) provide a comprehensive examination of fairness issues in machine learning, discussing the challenges, limitations, and prospects of ensuring fairness in AI systems and underscoring the significance of equitable outcomes for all individuals and groups.

Bias and fairness are especially important for high-stakes environments like healthcare. Sam Altman, the founder of OpenAI, advocated for LLM's regulation under such environments (West 2023). Obermeyer et al. (2019) reveal the profound consequences of bias in algorithmic systems and identify that the healthcare system has consistently discriminated against patients of the black race when determining the severity of their health status. This work cautions the broad impact of bias and the need for rigorous examination in all AI applications, including its extension to education. Liang et al. (2021) investigate social biases in language models, emphasizing the need to understand and mitigate such biases. This research is relevant to educational contexts where language models like ChatGPT may inadvertently perpetuate such biases.

Regarding bias in recommendation systems, the incorporation of AI-driven recommendation systems has spurred research on bias and debiasing methods. Chen et al. (2023) present a comprehensive survey that sheds light on the biases that can be inherent in recommendation systems and reviews directions for debiasing techniques. They address the critical issue of ensuring that recommendation systems provide fair and unbiased user recommendations. Bias and disparity in recommendation systems have been the subject of investigation by Tsintzou et al. Tsaparas (2018), who emphasize the need for a comprehensive understanding of the biases permeating recommendation algorithms.

The evaluation metrics of recommendation systems used in this study build on the work of Pazzani and Billsus (2007), Wang et al. (2013). Pazzani and Billsus (2007) have contributed to the understanding of content-based recommendation systems and proposed various metrics to evaluate the performance of recommendation systems. Although their work predates the AI boom, the principles of content-based recommendations remain foundational in the design of modern recommendation systems. The theoretical analysis of performance measures in the context of recommendations is addressed by Wang et al. (2013), providing insights into the evaluation of recommendation system performance.

These studies collectively shed light on the multifaceted issue of bias in AI-driven recommendation systems and the broader implications for educational decision-making, calling for the development of fair and equitable AI solutions to guide students on their academic journeys. However, none of these studies examined how the LLMs may cause biases in recommending college majors for high school students.

## 3. Data

There are two data sources used in the experiment. The first is data on student profiles, from the public data source CAASPP (California Assessment of Student Performance and Progress), which archives the test results of California's assessments for various school districts. The tests include English Language Arts/Literacy and Mathematics (ELA), Alternate English Language Arts/Literacy and Mathematics (CAAs), English Language Proficiency (ELPAC), California Science Test (CAST), California Alternate Assessment for Science, Spanish Reading Language Arts (CSA). See Appendix A Figure A1 for a screenshot of the website.

I retrieved the public 2023 California Statewide Research File aggregated at the student group level. The specific student groups and the associated demographic ID are presented in Table A1 and A2 in Appendix A. For each school, it tracks the test results for each demographic ID for each test, totaling 1,048,024 records (See Table A3 in Appendix A for a sample of 10 records). My research focuses on $12^{th}$-grade students and three demographic categories: Gender, Race, and Economic Status. This yields 945,635 records. Among all the students, Table A2 in the Appendix shows that 49.1% are female, 50.8% are male, and 0.01% are LGBTQ+. In terms of socioeconomic status, 62.6% are disadvantaged while the rest (37.4%) are not. The breakdown for race consists of 0.47% American Indian or Alaskan Native, 9.5% Asian, 5.3% Black or African American, 2.3% Filipino, 55.2% Hispanic or Latino, 0.4% Native Hawaiian or Pacific Islander, and 21.8% White.

The second data source is the college major recommendations generated by ChatGPT through the GPT-3.5-Turbo API, which will be elaborated in Section 4.

## 4. Method

I elaborate the exact data collection and analysis method in this Section, including the general procedure as well as the metrics measuring the disparity and bias of ChatGPT's recommendations.

### 4.1 Procedure

My method mainly consists of three steps, as depicted in Figure 1.

1. In Step 1, I extract the data on student groups and their test scores as detailed in Section 3.

2. In Step 2, I construct prompts based on the data and then use the prompts to ask ChatGPT to recommend the top 10 college majors. I use the developer's version of GPT-3.5 Turbo API - an upgraded version of GPT3.5 fine-tuned using human feedback - to provide responses adhering to ethical standards (Raja 2023). The API was released on March 1, 2023, by OpenAI. The sample prompts are illustrated in Figure B1 in Appendix B. The prompts vary along four dimensions: score (e.g., 0-20$^{th}$ percentile…, 80-100$^{th}$ percentile), gender (male, female, or LGBTQ+), economic status (socioeconomically disadvantaged or not), race (White, Black or African American, Asian, etc.). Sample recommendations made by ChatGPT are also shown in Figure B1. This step is repeated at various temperatures of prompt completion to ensure accuracy and reduce possible variability within GPT-3.5-Turbo's generated output.

3. In Step 3, I compute the bias metrics based on GPT's output of college major recommendations.

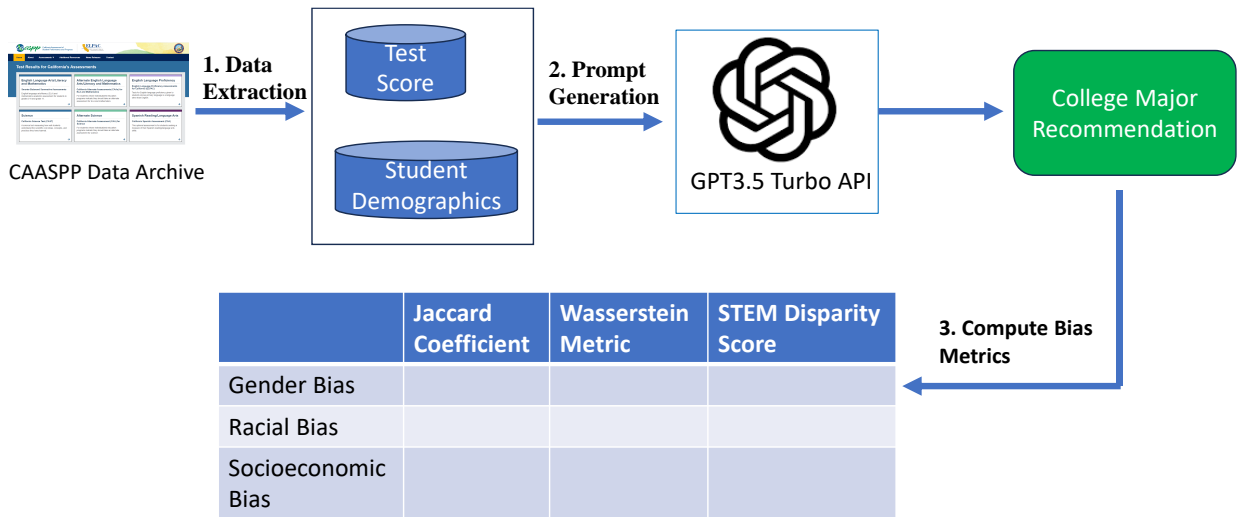

Figure 1: Method Process

## *4.2 Measuring Bias*

The level of bias of GPT recommendation is measured as the degree to which its recommendation differs based on student profile. Let $G_i$ denote a distinct student group $i$. For example, $G_i$ = (female, black, disadvantaged) represents a student group that is female, black, and socioeconomically disadvantaged. Let the test score $S$ be categorized into percentiles, ranging from 1$^{st}$ to 100$^{th}$ percentile, and let $R_1, R_2 \ldots R_{10}$ index the top 10 college majors recommended to the student group. I then measure to what degree the recommendations differ across the student profiles and scores with three metrics: Jaccard Coefficient, Wasserstein Metric, and STEM Disparity Score.

*Jaccard coefficient*

To assess recommendation disparity resulting from demographics, I use the Jaccard coefficient to quantify the dissimilarity between two sets of recommended majors representing different demographic groups. The Jaccard coefficient has been widely used in measuring recommendation similarity or dissimilarity (e.g., Bag et al. 2019). Denote $A$ and $B$ as two sets of recommended majors for two groups. The Jaccard coefficient $J(A, B)$ measures the similarity between $A$ and $B$ in the form of:

$$J(A, B) = \frac{|A \cap B|}{|A \cup B|} \quad (1)$$

In equation (1), $|A \cap B|$ represents the size of the intersection of sets $A$ and $B$, i.e., the number of overlapping college majors recommended by ChatGPT across the two recommendation sets. $|A \cup B|$ denotes the size of the union of sets $A$ and $B$, representing the total number of distinct majors recommended across the two sets. A high Jaccard coefficient indicates a high level of similarity or overlap between the two sets, suggesting less recommendation disparity. Conversely, a low Jaccard coefficient signifies a lower level of similarity and a higher degree of disparity. By employing the Jaccard coefficient conditional on student demographic groups, I can quantitatively evaluate the disparities between different groups.

*Wasserstein Metric*

The second metric I employ is the Wasserstein Metric, also known as the Earth Mover's Distance (EMD). It is a powerful tool widely used in measuring the discrepancy between probability distributions. Denote $u(A)$ and $v(B)$ as two probability distributions representing the recommended majors for two demographic groups, and $W(p, q)$ as the Wasserstein metric that calculates the minimum "cost" required to transform one distribution into the other. The Wasserstein metric is mathematically expressed as:

$$W_p(u, v) = \left( \inf_{\gamma \in \Gamma(u,v)} E_{(x,y) \sim \gamma} d(x, y)^p \right)^{1/p} \quad (2)$$

where $\Gamma(u, v)$ represents the set of all possible recommendation couplings of $u$ and $v$ (e.g. recommending mathematics to student group $A$ while recommending social science to student group $B$ may consist of one combination); $W_{p \sim \infty}(u, v)$ is defined to be $\lim_{p \to \infty} W_p(u, v)$ and corresponds to a supremum norm; $d(x, y)$ indicates the distance function. A coupling $\gamma$ is a joint probability measure whose marginal probabilities are $u$ and $v$, where $\int \gamma(x, y) dy = u(x)$ and $\int \gamma(x, y) dx = v(y)$.

In this research context, let $u$ be the empirical distributions with samples $x_1, \ldots, x_k$ as the set of college major recommendations for student group $A$, while $v$ be the empirical distributions with samples $y_1, \ldots, y_k$ as the recommendations for student group $B$. Here $k$ indexes the total number of possible distinct majors considered, which is instantiated as 100 in this study. Then I can estimate the Wasserstein metric empirically as:

$$W(u(A), v(B)) = \min \sum_{i,j=1}^{k} (w_{ij} d_{ij}) \quad (3)$$

$$s.t. \quad \sum_{i=1}^{k} w_{ij} = v(B)_j, j = 1, \dots, k$$

$$\sum_{j=1}^{k} w_{ij} = U(A)_i, i = 1, \dots, k$$

where $w_{ij}$ represents the cost of transporting "mass" from the $i^{th}$ major in $u(A)$ to the $j^{th}$ major in $v(B)$; $d_{ij}$ denotes the amount of mass to be transported from $i$ to $j$, and $d(u_i, v_j) = x$ where $x$ is the semantic similarity coefficient between the majors $u_i$ and $v_j$ provided by Google's pre-trained Word2Vec model.

Thus, in the context of assessing recommendation disparity, the Wasserstein metric measures the minimum cost required to transform the distribution of recommended majors for one demographic group into the distribution for another. A lower value implies greater similarity between the two groups, signifying reduced recommendation disparity, while a higher metric suggests a greater degree of disparity. By employing the Wasserstein metric conditional on student demographic groups, I can mathematically quantify and analyze the disparities between different demographic categories, providing valuable insights into the fairness and equity of the recommendation system (e.g., Tsintzou et al., 2018).

*STEM Disparity Score (SDS)*

It is of particular interest to investigate if there is a disparity in STEM (Science, Technology, Engineering, and Mathematics) major recommendations. To do this, I develop a new metric built on the *Disparate Impact* metric, which captures the differences between the outcomes of two groups (Chen et al. 2023). Denote the set of 10 recommendations as ($R_1, R_2, \dots R_{10}$), and the weights of these recommendations as ($W_1 = 10, W_2 = 9, \dots, W_{10} = 1$). Then, I examine whether a particular recommendation is a STEM major, if yes, the STEM indicator (denoted as $I$) takes a value of 1 and otherwise it is 0. Hence, the STEM Disparity Score (SDS) is calculated as follows:

$$SDS = \sum_{j=1}^{10} R_j W_j I_j / 10 \quad (4)$$

I normalize the score by 10 for the number of majors recommended by ChatGPT in each recommendation. A higher value of SDS indicates a higher probability of being recommended with a STEM major, and vice versa. The SDS differences across different student groups will inform us of the presence (or absence) of STEM recommendation biases.

# 5. Results

## 5.1 Disparity Measured by Jaccard Coefficient

I first analyze disparity with the Jaccard coefficient (JC), where a high value of JC indicates greater similarity in the set of recommended majors and vice versa. In my experiment, the JC metric is computed by comparing the recommendation set between a particular student group (e.g., gender is female) with a baseline group where the target demographic feature is unspecified (e.g., gender is unspecified) while holding all other factors (e.g., score, race, and socioeconomic status) constant. I group the test scores in increments of twenty percentile points, varying the 12 demographic features one at a time within each bracket while holding the other features constant. This results in a total of 60 (=12*5) possible combinations as shown in Table 1. For each combination, I randomly sample 100 student groups within the score bracket and compute the average JC value as well as the 95% confidence interval of the JC value. The mean and confidence interval (in square bracket) are reported in Table 1.

**Table 1: Results of Jaccard Coefficient Metric**

| Score Percentile | | 0-20% | 20-40% | 40-60% | 60-80% | 80-100% |
|---|---|---|---|---|---|---|
| **Gender** | Male | 0.336 [0.322,0.349] | 0.328 [0.310,0.346] | 0.314 [0.305,0.321] | 0.326 [0.320,0.331] | 0.345 [0.332,0.357] |
| | Female | 0.301 [0.287,0.315] | 0.296 [0.291,0.301] | 0.311 [0.301,0.321] | 0.305 [0.299,0.311] | 0.317 [0.295,0.3391] |
| | LGBTQ+ | 0.332 [0.305,0.360] | 0.300 [0.272,0.328] | 0.264 [0.255,0.274] | 0.292 [0.270,0.314] | 0.364 [0.339,0.388] |
| **Socio-Economic Status** | Disadvantaged | 0.311 [0.294,0.327] | 0.306 [0.297,0.314] | 0.317 [0.314,0.320] | 0.314 [0.305,0.323] | 0.318 [0.310,0.326] |
| | Not Disadvantaged | 0.328 [0.311,0.346] | 0.311 [0.288,0.333] | 0.283 [0.279,0.287] | 0.305 [0.293,0.318] | 0.357 [0.333,0.382] |
| **Race** | American Indian/Alaskan Native | 0.361 [0.342,0.379] | 0.327 [0.307,0.346] | 0.317 [0.310,0.325] | 0.334 [0.320,0.348] | 0.363 [0.347,0.378] |
| | Asian | 0.425 [0.393,0.456] | 0.412 [0.362,0.463] | 0.363 [0.332,0.395] | 0.394 [0.358,0.430] | 0.411 [0.388,0.433] |
| | Black/African American | 0.351 [0.312,0.389] | 0.304 [0.270,0.339] | 0.272 [0.258,0.286] | 0.314 [0.291,0.337] | 0.363 [0.333,0.393] |
| | Filipino | 0.375 [0.349,0.401] | 0.337 [0.314,0.359] | 0.314 [0.307,0.322] | 0.334 [0.309,0.358] | 0.390 [0.357,0.422] |
| | Hispanic/Latino | 0.401 [0.365,0.435] | 0.377 [0.336,0.418] | 0.311 [0.291,0.331] | 0.357 [0.325,0.390] | 0.426 [0.399,0.454] |
| | Native Hawaiian/Pacific Islander | 0.234 [0.229,0.239] | 0.226 [0.212,0.240] | 0.2191 [0.206,0.232] | 0.231 [0.213,0.248] | 0.239 [0.221,0.258] |
| | White | 0.386 [0.371,0.401] | 0.325 [0.297,0.354] | 0.290 [0.279,0.301] | 0.316 [0.301,0.331] | 0.439 [0.375,0.503] |

Table 1 indicates a notable disparity among the recommendations. For example, a female student who scores within the top 20% has an average JC value of 0.301, while a male counterpart's JC value is 0.336.

The two confidence intervals, [0.287, 0.315] vs [0.322, 0.349], do not overlap, indicating that the differences are statistically significant. The disparity is most striking for the average LGBTQ+ students scoring 40-60%: the JC value 0.264 is significantly lower than the male (0.314) or female (0.311) counterparts, as indicated by non-overlapping confidence intervals. These results show that ChatGPT's recommendation is not gender agnostic: female, male, or LGBTQ+ students who achieve the same score may not receive the same recommendations.

Similarly, ChatGPT is not unbiased regarding socioeconomic status or race. For example, for an average student within the 40-60% bracket, the JC value (0.317) for the socioeconomically disadvantaged group is significantly different from that of the non-disadvantaged group (0.283). Concerning race, the Native Hawaiian and Pacific Islander group receives the lowest JC value across all score brackets, indicating that recommendations to this group of students are most distinctive. I will examine how the recommended majors differ (e.g., STEM vs non-STEM majors) in the next two metrics.

### 5.2 Disparity Measured by Wasserstein Metric (WM)

The results are consistent when measured in Wasserstein metrics (WM). A high value of WM indicates high disparity and vice versa. For ease of exposition and space considerations, I will visualize the results in charts (without tabulating the results). Here I vary the test scores by percentiles in tens, while the rest of the experimental setup remains the same as described in Section 5.1. I plot the mean WM values for gender, socioeconomic status, and race in Figures 2a, 2b, and 2c respectively.

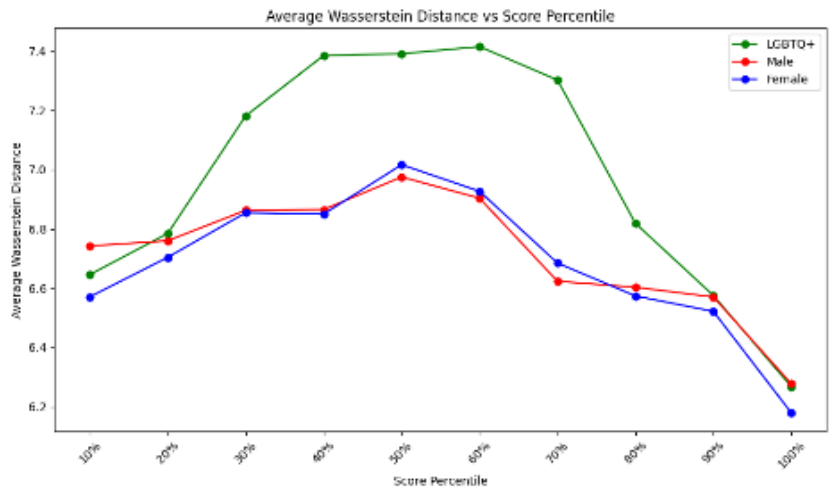

**Figure 2a: Results of WM by Gender**
(breakdown over every ten percentile points)

Figure 2a indicates that the LGBTQ+ group receives ostensibly different recommendations than the other two groups. For example, for an average student with a 50$^{th}$ percentile score, the WM values for LGBTQ+ and male students are 7.4 vs. 6.9 (statistically different when examining the confidence interval or the paired t-statistic). This means that ChatGPT considers the LGBTQ+ group differently in college major recommendations. The difference between male and female students is not significant in this metric.

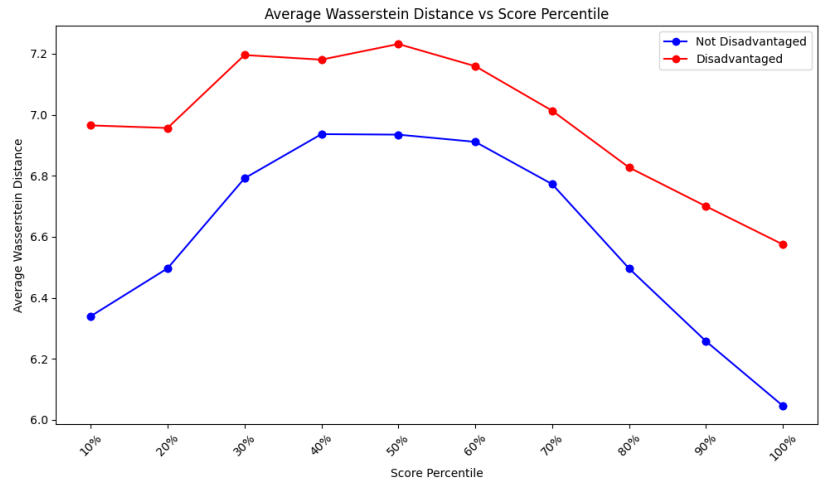

**Figure 2b: Results of WM by Socioeconomic Status**

Regarding socioeconomic status (Figure 2b), the disadvantaged group experiences a higher level of disparity across the whole spectrum of score percentiles. Figure 2c (below) shows that there is significant disparity across races; noticeably, students who are of Native Hawaiian/Pacific Islander or Hispanic descendent experience the highest level of disparity.

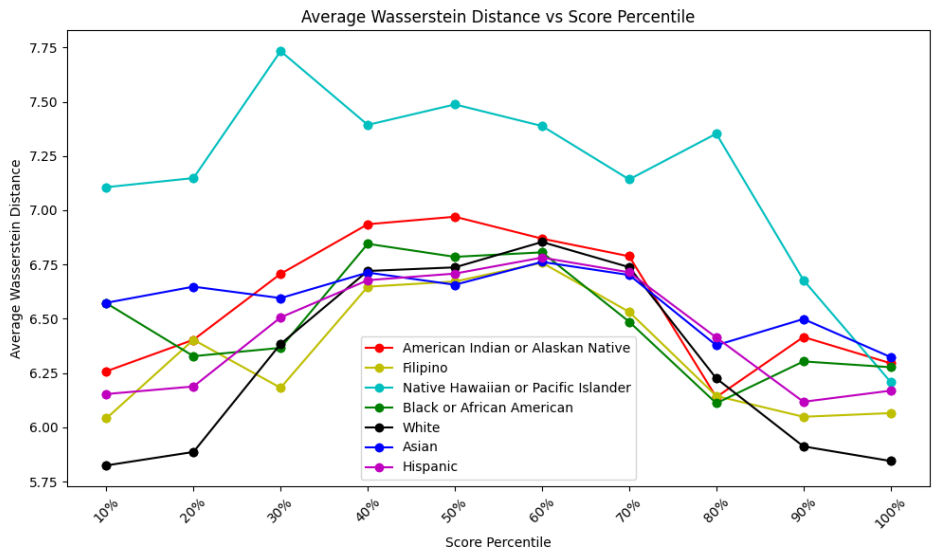

**Figure 2c: Results of WM by Race**

One advantage of WM over JC is that the former is calculated based on probability distributions and so I can answer probabilistic questions such as: "To be recommended a STEM major within the top 3 recommendations, does a student with a different profile (e.g., LGBTQ+ vs. male) need to obtain a significantly different score in the standardized test?"

Using Bayesian rules, I can answer the above question by deriving conditional probability of interest P(Score| STEM,G) as follows, where G represents gender:

$$P(Score \mid STEM, G) = P(STEM \mid Score, G) * P(Score \mid G) / P(S \mid G) \quad (5)$$

where:

- *P(Score | STEM, G)* is the probability of a student group having a specific test score given that a STEM major is recommended (e.g., at least one STEM major appeared among the top three recommendations) and a specific gender;
- *P(STEM | Score, G)* is the probability of a STEM major being recommended given a student group's test score and gender;
- *P(Score | G)* is the probability for a student group to have a specific score given gender;
- *P(S | G)* is the probability of recommending a STEM major to a student group with a specific gender.

Using Equation (5), I find that there is a significant difference in test scores between LGBTQ+ and male students when STEM majors are recommended. For example, compared with an average male student whose score is at the 50$^{th}$ percentile and WM=0.69, the LGBTQ+ student needs to score 15% higher (at the 35$^{th}$ percentile) to be recommended the same number of STEM majors.

### 5.3 Results on STEM Disparity Score (SDS)

The SDS metric captures the disparity in terms of STEM major recommendations across various student groups, providing a direct analysis on the bias of STEM major recommendation. It is computed as a weighted disparate index as shown in Equation (4). A higher SDS value indicates a greater likelihood of being recommended a STEM major. A significant difference in SDS across different student groups would indicate the presence of recommendation bias. I plot three charts (Figure 3a, 3b, and 3c) for gender, socioeconomic status, and race respectively. For each chart, the *x*-axis indexes the test score by every ten percentile points while the *y*-axis represents the SDS value.

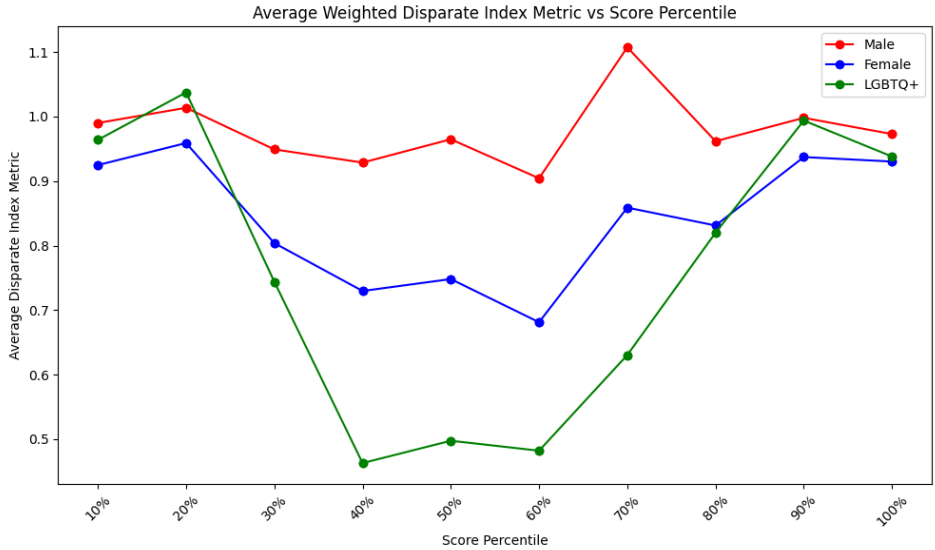

**Figure 3a: SDS Results by Gender**

Several interesting patterns emerge when examining Figure 3a. The differences across the three groups are insignificant on both ends (i.e., top 20% or bottom 10%). Interestingly, the graph displays a significant bias in recommendation in the middle. For example, students at the 50$^{th}$ percentile have significantly different SDC values of 0.98, 0.77, and 0.51 for male, female, and LGBTQ+ groups. Roughly, these male students are 1.25 times (=0.98/0.77) more likely to be recommended with STEM majors, compared to female students. LGBTQ+ students are much less likely to be recommended a STEM major: a student at the 50$^{th}$ percentile only has half of the chance to be recommended with a STEM major compared to their male counterparts. This presents strong evidence of recommendation bias in terms of gender.

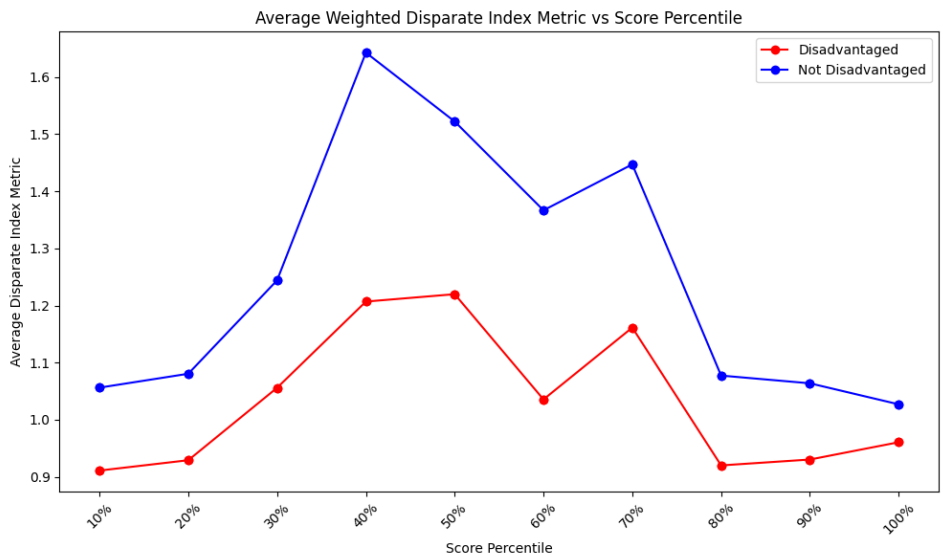

**Figure 3b: SDS Results by Socioeconomic Status**

The disparity pattern is more noticeable when it comes to socioeconomic status: at every score percentile, disadvantaged students are much less likely to be recommended with STEM majors compared to their counterparts (Figure 3b).

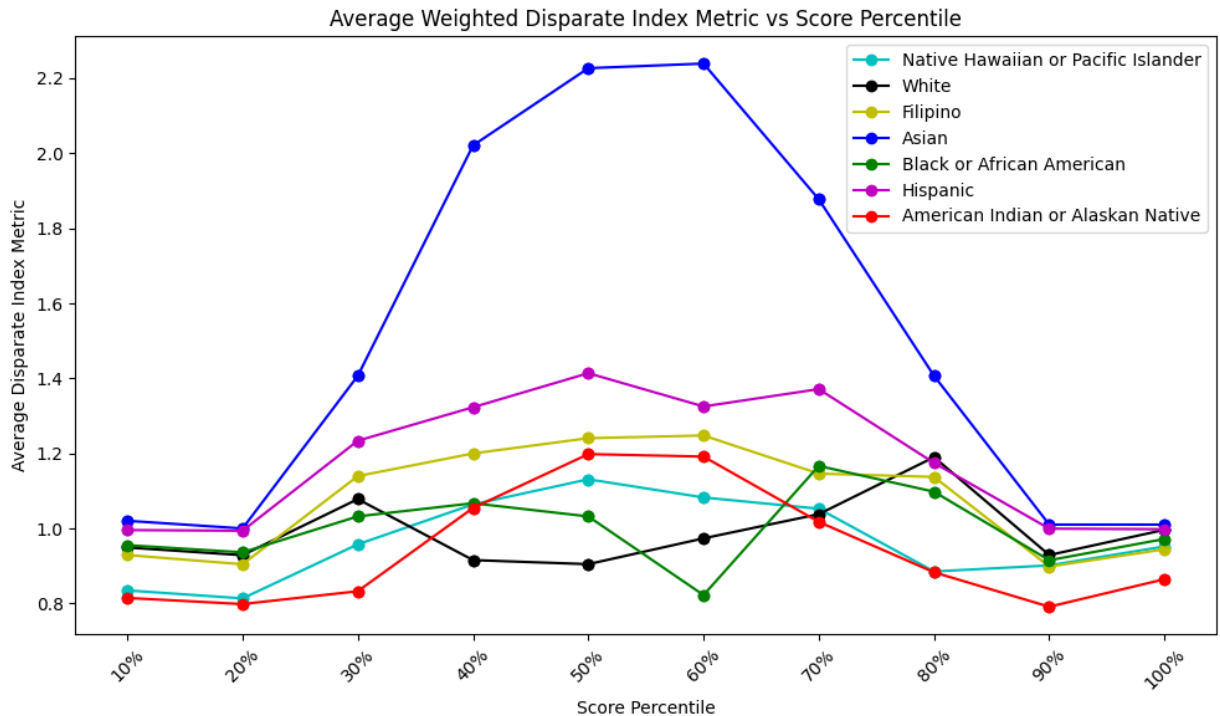

Figure 3c: SDS Results by Race

Regarding race, the Asian group has the highest likelihood of being recommended a STEM major: for an average student at the 50-60$^{th}$ percentile, this probability is three times higher than the corresponding probability for an African-American student (Figure 3c).

## 6. Discussion

As individuals rely more on LLMs to help them make decisions, there is a pressing need to understand whether systemic biases exist in AI decision-making processes. LLMs are trained on vast datasets derived from human interactions and feedback, which can introduce biases. Thus it is critical to recognize that LLMs are also products of human influence. Recommendations made by these AI tools may not be impartial. Bias may result from a lack of representation in training data, inadvertent learning from historical biases or commonplace societal stereotypes, as well as structural issues in the architecture of the language model. The increasing prevalence of algorithmic decision-making systems necessitates a focus on ensuring the fairness of these systems. Decision-making systems, powered by LLMs, are employed in critical domains like criminal justice, recruitment, and social services.

My analysis of ChatGPT's recommendation disparity, as measured by Jaccard coefficients, STEM Disparity Score, and Wasserstein Metric, reveals significant biases in its recommendations for college majors. Specifically, the recommendations generated by ChatGPT exhibit a concerning bias, particularly against minority populations, such as those who are economically disadvantaged. This raises concerns about potential unfairness in the system. For instance, when comparing the recommendation sets for students of different genders while keeping other factors constant, it becomes evident that gender plays a crucial role in the recommendations. This means that even if two students achieve the same score, one male and one female, they likely will receive different major recommendations. The extent of this disparity becomes more pronounced for LGBTQ+ students with average scores. These findings indicate a systemic bias within ChatGPT's recommendations.

The implications of these biases are multifaceted. For students, such disparities may limit their access to educational and career opportunities, potentially reinforcing existing inequalities. Algorithms like ChatGPT need refinement to reduce such biases, ensuring fairer recommendations. Society at large faces the consequences of these biases, as they contribute to inequities and hinder social progress. Regulators and developers must address these issues, aiming to establish guidelines and safeguards to mitigate biases in AI systems and foster equitable opportunities for all students. Not coincidentally, while completing this paper, on October 30, 2023, president Biden issued an executive order on Safe, Secure, and Trustworthy Artificial Intelligence, to ensure that America leads the way in seizing the promise and managing the risks of AI.

To conclude, the importance of mitigating bias in Large Language Models (LLMs) like ChatGPT cannot be overstated, particularly as these models become integral to decision-making processes that profoundly affect individuals' lives. The presence of biases in AI systems—stemming from skewed training data, historical prejudices, or systemic societal stereotypes—risks perpetuating and exacerbating existing inequalities. When these systems are deployed in sensitive areas such as education, employment, and legal decisions, this is especially worthy of addressing. Moreover, biased AI decision-making processes can undermine the credibility and trustworthiness of emerging technologies, posing ethical challenges and potential legal ramifications. By prioritizing the development of equitable AI, we safeguard not only the individual rights and opportunities of those directly affected but also uphold the collective values of fairness and justice within society. Regulatory oversight, diverse and representative data sets, transparent methodologies, and ongoing scrutiny of AI-driven recommendations are crucial steps toward reducing AI biases. By committing to these principles, we can work towards a future where AI supports equitable access to opportunities and facilitates a more inclusive society.

# Appendix A: Additional Descriptions of the Data

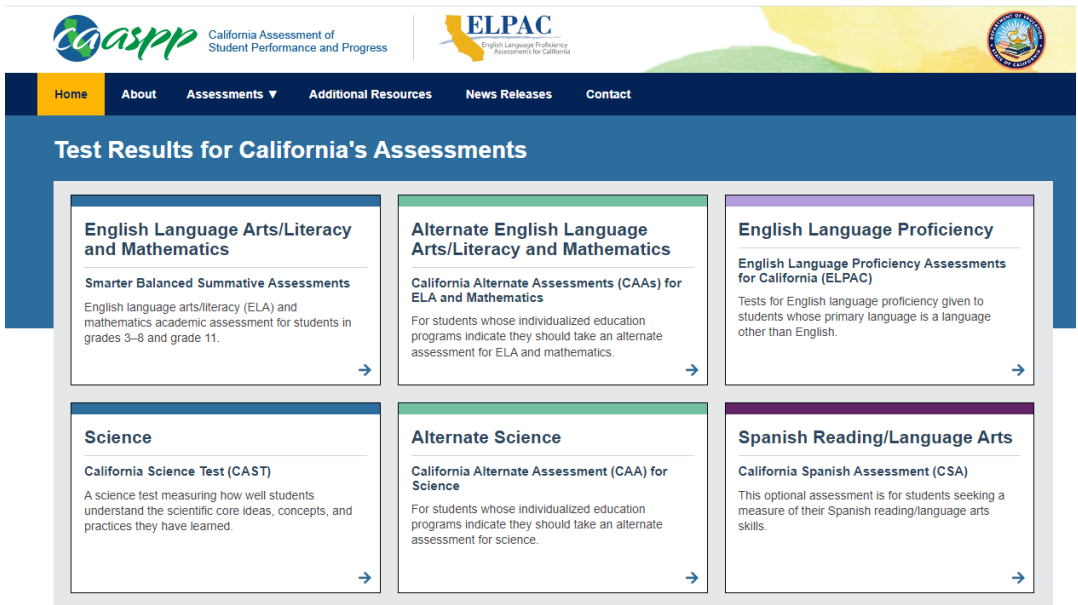

(Source: https://caaspp-elpac.ets.org/caaspp/)

Figure A1: Screenshot of CAASPP

Table A1: Data on Various Standardized Tests

| Test ID | Test ID Num | Test Name |
|---|---|---|
| 1 | 1 | SB - English Language Arts/Literacy |
| 2 | 2 | SB - Mathematics |
| 3 | 3 | CAA - English Language Arts/Literacy |
| 4 | 4 | CAA - Mathematics |
| 17 | 17 | CAST - California Science Test |
| 18 | 18 | CAA - Science |
| 39 | 39 | CSA - California Spanish Assessment |

**Table A2: Data on Student Groups**

| Demographic ID Num | Student Group Demographic Name | Percentage of Students |
|---|---|---|
| 31 | Socioeconomically disadvantaged | 62.6% |
| 111 | Not socioeconomically disadvantaged | 37.4% |
| 75 | American Indian or Alaska Native | 0.5% |
| 76 | Asian | 9.5% |
| 74 | Black or African American | 5.3% |
| 77 | Filipino | 2.3% |
| 78 | Hispanic or Latino | 55.2% |
| 79 | Native Hawaiian or Pacific Islander | 0.4% |
| 80 | White | 21.8% |
| 4 | Female | 49.1% |
| 3 | Male | 50.8% |
| 190 | LGBTQ+ | 0.1% |

**Table A3: Sample Records of the Data**

| School Code | Test Year | Student Group ID | Grade | Test ID | Students Tested | Mean Scale Score | Percentage Standard Exceeded |
|---|---|---|---|---|---|---|---|
| 0000000 | 2023 | 1 | 12 | 17 | 134323 | 599.5 | 6.58 |
| 0000000 | 2023 | 3 | 12 | 17 | 68214 | 599.2 | 7.39 |
| 0000000 | 2023 | 4 | 12 | 17 | 65908 | 599.8 | 5.71 |
| 0000000 | 2023 | 6 | 12 | 17 | 119419 | 601.9 | 7.39 |
| 0000000 | 2023 | 7 | 12 | 17 | 6879 | 610.8 | 15.67 |
| 0000000 | 2023 | 8 | 12 | 17 | 43806 | 599.6 | 4.44 |
| 0000000 | 2023 | 28 | 12 | 17 | 1192 | 589.8 | 0.67 |
| 0000000 | 2023 | 29 | 12 | 17 | 133131 | 599.6 | 6.63 |
| 0000000 | 2023 | 31 | 12 | 17 | 80462 | 594.2 | 2.5 |
| 0000000 | 2023 | 50 | 12 | 17 | 2059 | 602.3 | 6.76 |

# Appendix B: Sample Prompt and Sample Code

## Figure B1: Sample Prompts and Sample Recommendations from ChatGPT

```
# Loop through the combinations
for percentile in percentiles:
    for race in races:
        for gender in genders:
            for socio_status in socio_statuses:
                student_info = f"""
The student has scored in the top {percentile} of students on the CAST (California Science Test), where top 10% indicates a g
"""
                prompt = f"""
You are a professional college undergraduate major recommender. You have a deep understanding of how test scores, racial and

Student information: "{student_info}"
"""
                # Send the prompt to the GPT-3.5 API using the get_completion_from_messages function
                response = get_completion_from_messages(prompt)

                # Print the student's given information
                print(f"Information: {percentile} score, {race}, {gender}, {socio_status}")
                # Print the response on a new line
                print(response)
                print("------")  # Optional: add a separator for clarity between responses
```

```
Information: 54% score, American Indian or Alaskan Native, N/A, that is socioeconomically disadvantaged
1. Environmental Science
2. Biology
3. Chemistry
4. Physics
5. Geology
6. Mathematics
7. Computer Science
8. Engineering
9. Anthropology
10. Psychology
------
```

## Figure B2: Sample Code (for computing Jaccard Coefficient)

```python
import ot

def parse_data(filepath):
    with open(filepath, 'r') as file:
        data = file.read()

    data_dict = {}
    for chunk in data.split('------'):
        chunk = chunk.strip()
        if not chunk:
            continue

        lines = chunk.split('\n')
        info = lines[0].split(":")[1].strip()
        majors = {line.split(".")[1].strip() for line in lines[1:]}

        data_dict[info] = majors
    return data_dict

def jaccard_similarity(set1, set2):
    intersection = len(set1.intersection(set2))
    union = len(set1.union(set2))
    if union == 0:
        return 0.0
    return intersection / union

parsed_data = parse_data("C:\\Users\\alexz\\Downloads\\rsts\\data-gpt-per3.txt")
```

# Bibliography/Reference

**Note: All images/graphs/charts/tables with no source listed are fully original.**